\documentclass[runningheads]{llncs}
\usepackage[T1]{fontenc}
\usepackage[misc]{ifsym}
\usepackage{gensymb}
\usepackage{graphicx}
\usepackage{array}
\usepackage{cite}
\usepackage{amsmath,amssymb,amsfonts}
\usepackage{algorithmic}
\usepackage{graphicx}
\usepackage{textcomp}
\usepackage{multirow}
\usepackage{booktabs}
\usepackage{url}
\usepackage{tablefootnote}
\usepackage{stfloats}
\usepackage{color}
\usepackage{verbatim}
\usepackage{bigstrut}
\usepackage{makecell}
\usepackage{caption}
\usepackage[marginal]{footmisc}

\usepackage{float}
\begin{document}
\begin{sloppypar}
\title{Hierarchical Characterization of Brain Dynamics via State Space-based Vector Quantization
}
  \author{Yanwu Yang\inst{1} \and
  Thomas Wolfers\inst{1,2}$^{(\textrm{\Letter})}$}
  \authorrunning{Yanwu, Thomas.}
  \institute{University of Tübingen, Tübingen, Germany  \and
  German Center for Mental Health, Tübingen, Germany \\
  \email{yangyanwu1111@gmail.com }
  \email{dr.thomas.wolfers@gmail.com }
  }

\maketitle              
\begin{abstract}
Understanding brain dynamics through functional Magnetic Resonance Imaging (fMRI) remains a fundamental challenge in neuroscience, particularly in capturing how the brain transitions between various functional states. Recently, metastability, which refers to temporarily stable brain states, has offered a promising paradigm to quantify complex brain signals into interpretable, discretized representations. In particular, compared to cluster-based machine learning approaches, tokenization approaches leveraging vector quantization have shown promise in representation learning with powerful reconstruction and predictive capabilities. However, most existing methods ignore brain transition dependencies and lack a quantification of brain dynamics into representative and stable embeddings. In this study, we propose a Hierarchical State space-based Tokenization network, termed HST, which quantizes brain states and transitions in a hierarchical structure based on a state space-based model. We introduce a refined clustered Vector-Quantization Variational AutoEncoder (VQ-VAE) that incorporates quantization error feedback and clustering to improve quantization performance while facilitating metastability with representative and stable token representations. We validate our HST on two public fMRI datasets, demonstrating its effectiveness in quantifying the hierarchical dynamics of the brain and its potential in disease diagnosis and reconstruction performance. Our method offers a promising framework for the characterization of brain dynamics, facilitating the analysis of metastability.

\end{abstract}
\section{Introduction}
In recent years, functional magnetic resonance imaging (fMRI), which uses the blood oxygen level-dependent (BOLD) effect, has emerged as a powerful tool to study temporal brain dynamics in cognitive processes and characterize dysfunction in patients \cite{yangMappingMultimodalBrain2023, jeong2024brainwavenet,yang2024brainmass}. Despite advances in modeling paradigms \cite{song2025bpclipbottomupimagequality,yang2023creg,yang2024advancing}, understanding these dynamic patterns remains a challenge in neuroscience, as the human brain exhibits dynamic high-dimensional, non-linear, and rapidly evolving dynamics that continuously shift between distinct functional states in response to varying cognitive demands \cite{beim2019metastable, cavanna2018dynamic}.
Recent studies on brain metastability have provided a new perspective for understanding these complex dynamics \cite{rabinovich2008transient}. Metastability refers to temporarily stable brain states that persist before transitioning to different configurations, offering a paradigm to quantify continuous brain signals into discretized, and interpretable representations. This approach not only captures the inherent temporal discontinuities of neural activity but also provides a principled way to quantize continuous brain dynamics into finite meaningful states \cite{rabinovich2008transient}.


However, challenges remain in quantifying brain dynamics through the lens of metastability. Traditional approaches, such as LEiDA \cite{cabral2017cognitive,kurtin2023task}, rely on simple clustering methods to identify patterns of recurrent brain function. 
However, their linearity and reliance on statistical clustering may oversimplify brain dynamics by ignoring nonlinear interactions, failing to capture rapid state transitions, and reducing high-dimensional data to overly simplistic partitions.
Recent advances in deep learning have introduced tokenization or quantization approaches as a promising solution \cite{van2017neural,song2025LVPNet}. These methods map inputs into structured representations such as discrete tokens. Vector Quantized Variational Autoencoders (VQ-VAE) \cite{van2017neural} have demonstrated remarkable success in various domains, enabling probabilistic modeling and token prediction for pre-training \cite{devlin2018bert, baevski2020wav2vec, bao2021beit}. Unlike traditional VAEs that learn continuous latent spaces, VQ-based methods map features to a finite codebook of embeddings, producing more compact and interpretable representations. Several variants have been proposed to improve the reconstruction and quantization performance, including VQ-VAE-2, Online Cluster VQ-VAE, and RQ-VAE \cite{razavi2019generating, zheng2023online, lee2022autoregressive}.

Despite advances in vector quantization, most existing approaches on brain dynamics potentially ignore brain state transitions in neural dynamics, which underlie the mechanisms of cognitive flexibility and dynamic patterns of neural activity between different cognitive states or processing modes. Studies have shown that abnormal transition patterns are associated with various neurological and psychiatric conditions, making them crucial for understanding brain function \cite{liu2019tracking,du2021abnormal}. Moreover, another significant challenge lies in the trade-off between quantization precision and stability \cite{zhao2024representation,zheng2023online}.
Increasing the size of the codebook tends to improve reconstruction performance, but it can lead to unstable and fragmented token representations that are sensitive to noise in neural signals \cite{zhao2024representation,razavi2019generating}. This instability makes it difficult to characterize the metastability in brain dynamics under different experimental conditions or between individuals.

In this paper, we propose to investigate the characterization of brain dynamics with a hierarchical structure that captures both brain states and transitions simultaneously. Moreover, to improve quantization stability and reconstruction performance, we introduce a refined cluster-based VQ-VAE by optimizing codebook utilization and clustering effectiveness. Our key contributions are:

(1) We propose a state space-based tokenization method for the quantization of time series fMRI data that enables hierarchical mapping of brain states and transitions.

(2) We introduce a refined cluster VQ-VAE that incorporates quantization error feedback and clustering to improve quantization performance while facilitating metastability with stable and representative discretized embeddings.

(3) Evaluations on two public datasets for disease diagnosis and reconstruction performance highlight the superiority of our method in brain dynamics quantization and its potential for analyzing clinical brain states.

\section{Method}
\begin{figure}[t]
    \centering
    \includegraphics[width=1.0\textwidth ]{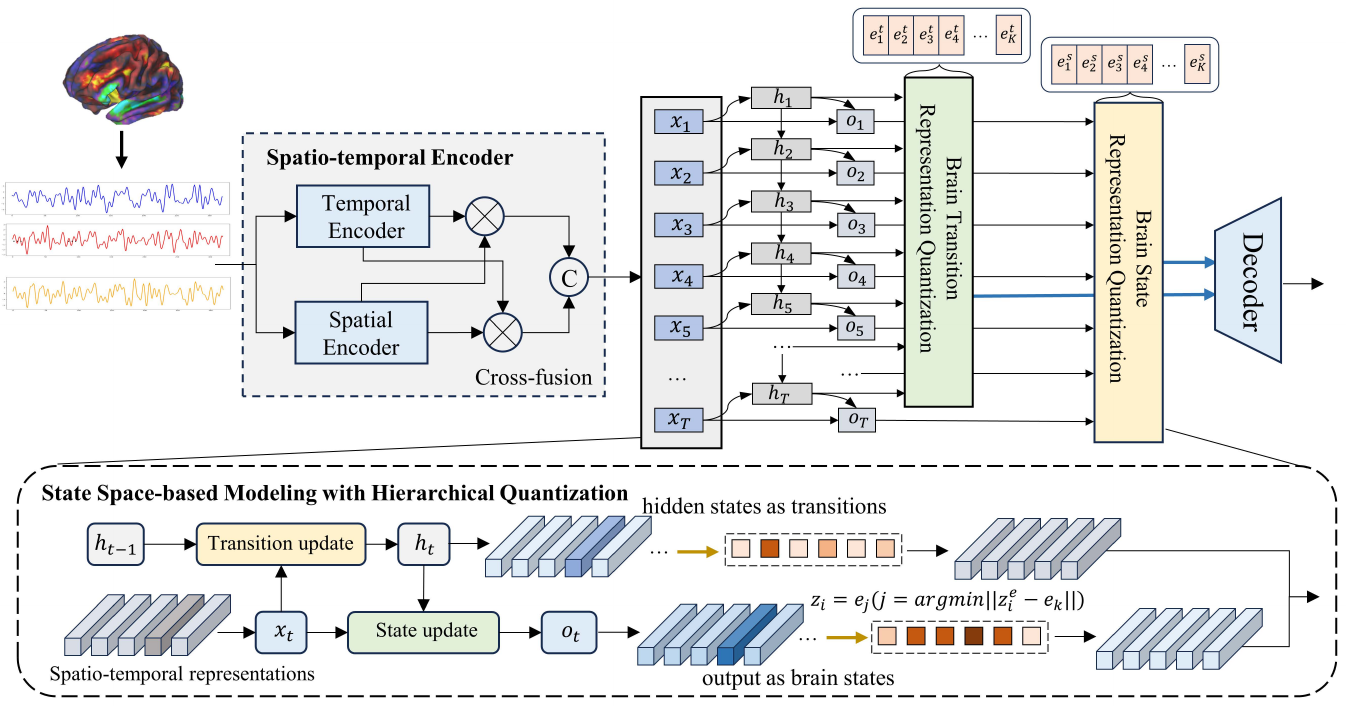}
    \caption{
    The architecture of our HST, including a spatio-temporal encoder, and a state space-based model with a hierarchical VQ-VAE for quantization. For downstream tasks, the quantization embeddings are concatenated and fed into a multi-layer perceptron for classification.}
    \label{model}
\end{figure}

The framework of our proposed HST is shown in Fig. \ref{model}. The inputs are first encoded by a spatio-temporal encoder and then fed into a hierarchical vector quantization based on brain state and transition representations. The quantized embeddings are concatenated and decoded for reconstruction.

\subsection{Spatio-temporal Encoder}
For a given input BOLD signal $X \in \mathbb{R}^{T \times M}$, the spatio-temporal encoder encodes the sequence from both temporal and spatial perspectives and then fuses them attentively. The temporal and spatial encoding modules are implemented using transformer networks as:
$H^{t} = \text{Enc}(X^{\text{temporal}})$, where $X^{\text{temporal}} = {X_{1,:}, X_{2,:}, ..., X_{T,:}}$, and $H^{s} = \text{Enc}(X^{\text{spatial}})$, where $X^{\text{spatial}} = {X_{:,1}, X_{:,2}, ..., X_{:,M}}$. In particular, $X^{\text{temporal}}$ and $X^{\text{spatial}}$ are obtained by sequencing the BOLD signal along the temporal and spatial dimensions. For each encoder ($\text{Enc}$), $H = \text{MHSA}(X)$, based on the query, key, and value matrices $Q_t^{l,c}, K_t^{l,c}, V_t^{l,c}$ and $Q_s^{l,c}, K_s^{l,c}, V_s^{l,c}$. The temporal and spatial attention scores are obtained by squeezing and exciting the corresponding representations:
$A^t = \sigma(w^s_2 \cdot \delta(w^s_1 \cdot \frac{1}{M}\sum_{j=1}^M H^t_{:,j} + b^s_1) + b^s_2)$, where $A^t \in \mathbb{R}^{T \times 1}$, and $A^s = \sigma(w^t_2 \cdot \delta(w^t_1 \cdot \frac{1}{T}\sum_{i=1}^T H^t_{i,:} + b^t_1) + b^t_2)$, where $A^s \in \mathbb{R}^{1 \times M}$
Here, $w^t_1, w^t_2, w^s_1, w^s_2$ and $b^t_1, b^t_2, b^s_1, b^s_2$ are learnable parameters for attention projection. Finally, cross-matrix multiplication is used for fusion of spatial and temporal representations: $H^{f} = H^t \otimes A^s + H^s \otimes A^t$.


\subsection{State Space-based Modeling Encoder}
The cross-fusion spatio-temporal representations are split into sequences $H^f = {H^f_1, H^f_2, ..., H^f_T}$ along the temporal dimension. These sequences are then fed into the state space-based modeling encoder for encoding brain dynamic states and transitions between states. In general, this is obtained as:
\begin{equation}
h_t = f(h_{t-1}, x_t, ...; \Theta), \quad
o_t = g(h_t, x_t, ...; \Phi)
\end{equation}
Here, $f$ and $g$ denote the mapping functions for transition $h_t$ and state $o_t$ updates at the $t$-th step, with $\Theta$ and $\Phi$ as their corresponding parameters. In particular, for each update, the $t$-th transition and state representations depend on previous representations $h_{t-1}$, or $h_{t-2}$, varying with the dependency range of different methods. In our study, we examine this using various approaches: RNN updating on the previous step, GRU and LSTM with attentive forgetting and updating, and Mamba with selective state updates.
For RNN, the state updates as follow:
\begin{equation}
h_t = \sigma(W_h h_{t-1} + W_x H_{t} + b_h)
\end{equation}
For Mamba \cite{gu2023mamba}, the states are obtained by:
\begin{gather}
\overline{\mathbf{A}} = \exp (\mathbf{\Delta A}),\quad
\overline{\mathbf{B}} = ( \mathbf{\Delta A})^{-1}(\exp(\mathbf{\Delta A})- \mathbf{I})\cdot \mathbf{\Delta B},\quad
h_t = \overline{\mathbf{A}} h_{t-1}+\overline{\mathbf{B}} x_t
\end{gather}
where $\overline{\mathbf{A}}$ and $\overline{\mathbf{B}}$ are discrete equivalents of the corresponding continuous evolution matrix $\mathbf{A}$ and projection matrix $\mathbf{B}$. $\mathbf{\Delta A}$ and $\mathbf{\Delta B}$ are learnable parameters representing the discretized versions of the continuous state space-based model, where $\mathbf{\Delta A}$ captures the state transitions over time steps. More details can be found in the previous paper \cite{gu2023mamba}. We formulate the state output representations using a two-layer MLP as:
$
o_t = W_{o2} (W_{o1} (h_{t-1} \parallel x_t) + b_{o1}) + b_{o2}
$.

\subsection{Refined Cluster-based VQ-VAE}
While deep learning models often benefit from larger codebooks, metastability requires fewer and more meaningful clusters. We address this by proposing a refined cluster VQ-VAE that optimizes codebook efficiency, achieving high reconstruction quality with a compact, interpretable codebook.
Our model employs a hierarchical architecture with error feedback and dynamic clustering. The first level implements separate quantization for brain states and transitions:
\begin{equation}
\hat{o}_t = e_k^S(k=\text{argmin}_{i} ||o_t - e_i^S||_2), \quad
\hat{h}_t = e_k^T(k=\text{argmin}_{j} ||h_t - e_j^T||_2)
\end{equation}
where $e^S$ and $e^T$ denote the embeddings of brain state and transition codebooks, and $\hat{o}$ and $\hat{h}$ are the corresponding representations after quantization.
Secondly, an auxiliary quantization module computes and feeds back errors to optimize the codebooks:
\begin{gather}
\hat{\epsilon}_{o,t}= e_{o,k}^S(k=\text{argmin}_{p}||(o_t-\hat{o}_t)-e^S_p||_2), \\
\hat{\epsilon}_{h,t}= e_{h,k}^S(k=\text{argmin}_{q}||(h_t-\hat{h}_t)-e^T_q||_2)
\end{gather}
where $e^S_{o}$ and $e^T_{h}$ are the corresponding error feedback codes.
Finally, the clustering quantization dynamically updates underutilized code vectors by repositioning them closer to active feature regions:
\begin{equation}
\mathbf{e}_j^{(t+1)} = \mathbf{e}_j^{(t)} \cdot (1 - \alpha_j) + \hat{\mathbf{z}}_j \cdot \alpha_j
\end{equation}
where $\alpha_j = \exp(-N_j \cdot K \cdot \frac{10}{1 - \gamma})$, and $N_j^{(t+1)} = \gamma N_j^{(t)} + (1 - \gamma) n_j^{(t)}$ \cite{zheng2023online}. Here, $\hat{\mathbf{z}}_j$ represents the feature closest to the $j$ -th codebook vector.

\subsection{Decoder and Optimization}
After quantifying the spatio-temporal dynamic representations of the brain into state and transition tokens, we leverage a Transformer network to decode the latent embeddings into BOLD signals. The codebook embeddings of brain states and transitions are concatenated and then fed into the decoder. Finally, the loss function is to optimize the reconstruction more closer to the input, as well as a weighted combination of codebook loss with $\alpha$ and $\beta$. We use $\text{sg}(\cdot)$ as a gradient stop operation to control the flow of gradients:
\begin{equation}
\begin{aligned} 
    \mathcal{L} = \alpha \|x-\hat{x}\|_2^2 &+ \beta\|\text{sg}(o_t) - \hat{o}_t\|_2^2 + \gamma\|\text{sg}(o_t - \hat{o}_t) - e_t\|_2^2 \\
    &+ \beta\|\text{sg}(h_t) - \hat{h}_t\|_2^2 + \gamma\|\text{sg}(h_t - \hat{h}_t) - e_t\|_2^2
\end{aligned}
\end{equation}

\section{Experiments and results}
\subsection{Dataset and experimental settings}
\textbf{Datasets.} In this study, we use two fMRI datasets on different scales for evaluation. (1) \textbf{The ADHD-200 dataset} \cite{adhd2012adhd} is a publicly available collection of 1,260  individuals in multiple data centers. After excluding images with scan durations shorter than 100 TRs, our final sample includes 954 subjects: 405 patients with Attention Deficit Hyperactivity Disorder (ADHD) and 549 healthy controls (HC) with sex and age matching. (2) \textbf{The SchizoConnect dataset} \cite{wang2016schizconnect} is a multisite neuroimaging resource providing MRI data in schizophrenia studies, comprising 97 patients with schizophrenia (SCZ) and 91 healthy controls (HC).

\textbf{Pre-processing.} All images were pre-processed using Configurable Pipeline for the Analysis of Connectomes (C-PAC) software, including anterior commissure-posterior commissure (AC-PC) alignment, skull stripping, bias field correction, and normalization into the standard Montreal Neurological Institute (MNI) space.

\textbf{Implementations}.
The transformer networks in the spatio-temporal encoder and decoder are implemented with 2 layers and 4 heads. The spatial state modeling networks are analyzed in the results section, including RNN, LSTM, GRU, and Mamba, where the hidden feature size is set to 256 and the number of layers is 2. The number of tokens in brain state and transition quantization is searched within the range $[8, 16, 32, 64, 128]$.
For hyperparameters, $\beta$ is set to $0.1$, and $\alpha$ is set to $1.0$. The learning rate is set to $0.0002$. All models are trained using the Adam optimizer. The training process consists of two phases. The tokenizer quantization training is conducted for 4,000 steps, followed by 200 epochs of downstream evaluation. During the second phase, the quantizers are frozen and only the encoder and a 3-layer MLP classifier are trained for classification.


\begin{figure}[t]
  \centering
  \includegraphics[width=0.95\textwidth ]{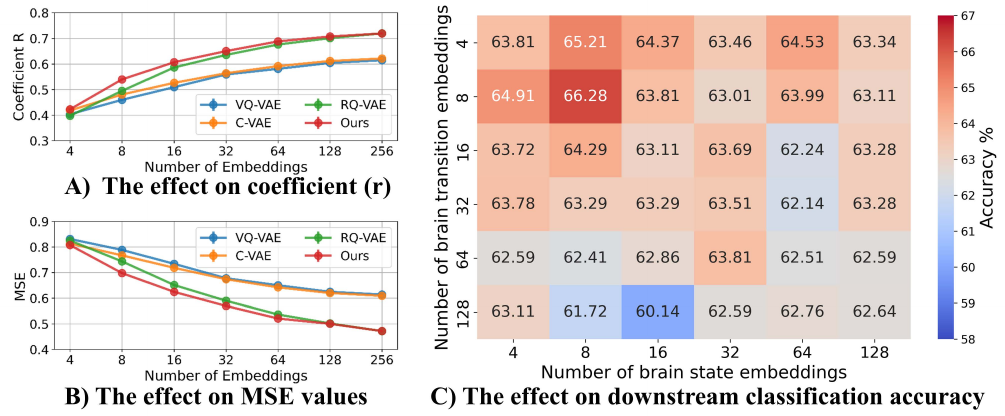}
  \caption{
  Reconstruction performance on the test set of ADHD-200 dataset in terms of A) correlation (r) and B) Mean Squared Error (MSE), and C) downstream classification accuracy (Acc) with different numbers of embeddings.}
  \label{ablation}
\end{figure}

\begin{table}[htbp]
    \centering
    \scriptsize
      \renewcommand\arraystretch{1.1}
    \setlength{\tabcolsep}{4pt}
    \caption{Diagnosis performance based on different state-space modeling methods.}
  \begin{tabular}{c|ccc|ccc}
  \hline
  \hline
  & \multicolumn{3}{c|}{ADHD-200} & \multicolumn{3}{c}{SchizoConnect} \\
  \cline{2-7}      & Acc   & Sen   & Spe   & Acc   & Sen   & Spe \\
  \hline
  RNN   & 63.81$\pm$2.98 & 61.81$\pm$5.42 & 64.91$\pm$1.74 & 71.57$\pm$2.52 & 68.45$\pm$3.57 & 77.94$\pm$3.08\\
  LSTM  & 65.03$\pm$3.01 & 64.10$\pm$3.77 & 65.78$\pm$3.40 & 73.71$\pm$5.05 & 70.36$\pm$4.82 & 79.18$\pm$6.43 \\
  GRU   & 62.59$\pm$2.30 & 59.53$\pm$4.65 & 65.03$\pm$3.35 & 73.57$\pm$2.52 & 69.45$\pm$3.57 & 80.94$\pm$3.08 \\
  Mamba & \textbf{66.28$\pm$2.54} & \textbf{66.68$\pm$3.73} & \textbf{66.18$\pm$3.87} & \textbf{75.72$\pm$2.15} & \textbf{75.00$\pm$3.00} & \textbf{83.75$\pm$4.25}\\
  \hline
  \hline
  \end{tabular}%
    \label{res_rnn}%
  \end{table}%

\subsection{Results}
\textbf{1. The effect of the number of quantization embeddings.}
We first evaluated the effect of embedding numbers on reconstruction and downstream classification. The state space-based model was implemented using Mamba \cite{gu2023mamba}. We show the results of the ADHD-200 dataset in Figure \ref{ablation}. Figure \ref{ablation} A) and B) demonstrate that our method achieved better reconstruction performance (higher correlation and lower MSE) with fewer embeddings, compared with other baselines. In particular, HST with 8 embeddings even outperformed both C-VAE and VQ-VAE using 16 embeddings. This demonstrates the efficiency of using a hierarchical quantization structure.
For downstream classification, optimal performance was achieved using 4-16 brain state embeddings and transitions, as shown in Figure \ref{ablation} C). Although increasing the size of the codebook improved reconstruction, the classification performance initially increased and then decreased. This reveals a trade-off between reconstruction fidelity and discriminative power. In general, using eight embeddings for quantifying brain states and transitions yields the best classification performance. Similarly, a grid search on the SchizoConnect dataset showed that 16 embeddings provided the best results.

\begin{table}[htbp]
    \centering
    \scriptsize
    \setlength{\tabcolsep}{1.5pt}
    \caption{Classification results based on different sequential modeling methods. The best results are shown in bold.}
\begin{tabular}{c|ccc|ccc}
  \hline
  \hline
        & \multicolumn{3}{c|}{ADHD-200} & \multicolumn{3}{c}{SchizoConnect} \\
  \cline{2-7}      & Acc   & Sen   & Spe   & Acc   & Sen   & Spe \\
  \hline
  SVM   & 57.16$\pm$3.14 & 56.08$\pm$4.64  & 57.84$\pm$3.89 & 59.46$\pm$2.96 & 57.39$\pm$1.60 & 58.87$\pm$3.89 \bigstrut[t]\\
  MLP   & 59.60$\pm$2.40 & 55.21$\pm$5.41 & 59.46$\pm$2.52 & 64.61$\pm$5.05 & 68.86$\pm$1.36 & 68.53$\pm$8.27 \\
  ST-GCN \cite{gadgil2020spatio} & 60.73$\pm$2.24 & 60.17$\pm$1.88 & 59.63$\pm$2.21 & 66.25$\pm$1.54 & 65.87$\pm$2.99 & 68.90$\pm$4.51 \\
  FBNetGen \cite{kan2022fbnetgen} & 61.41$\pm$2.54 & 62.87$\pm$2.00 & 60.08$\pm$2.67 & 71.97$\pm$1.54 & 70.96$\pm$4.64 & 73.90$\pm$2.63 \\
  BolT \cite{bedel2023bolt} & 63.51$\pm$2.55 & 59.28$\pm$4.39 & 63.17$\pm$3.12 & 71.22$\pm$1.60 & 74.45$\pm$2.52 & 70.60$\pm$2.04 \\
  BrainWaveNet \cite{jeong2024brainwavenet} & 64.95$\pm$2.92 & 65.39$\pm$3.30 & 65.99$\pm$5.44 & 73.75$\pm$3.89 & 73.12$\pm$3.91 & 75.61$\pm$6.68 \\
  HST (Ours) & \textbf{66.28$\pm$2.54} & \textbf{66.68$\pm$3.73} & \textbf{66.18$\pm$3.87} & \textbf{75.72$\pm$2.15} & \textbf{75.00$\pm$3.00} & \textbf{83.75$\pm$4.25}\\
  \hline
  \hline
  \end{tabular}%
    \label{res_bench}%
  \end{table}%

\textbf{2. The effect of different SSM methods on diagnostic predictions.}
We further evaluated four state space-based modeling methods (RNN, LSTM, GRU, and Mamba) for ADHD/SCZ diagnosis across both datasets. Table \ref{res_rnn} presents the performance metrics of accuracy, sensitivity, and specificity across folds, reported as the mean and standard deviation, with the best results highlighted in bold. The results show that Mamba consistently outperformed the other approaches, highlighting the effectiveness of its selective long-range modeling capabilities. Although other methods achieved comparable performance in some cases, Mamba's consistently superior performance across metrics and datasets underscores its robustness in sequential modeling tasks.

\textbf{3. Comparisons with other baselines on diagnostic predictions.}
Furthermore, we compared our HST with other sequential modeling methods for disease diagnosis. For downstream classification tasks, we extracted the hidden embeddings from the layer preceding the decoder and passed them through an MLP classifier. During this phase, we kept the codebook parameters fixed while fine-tuning both the encoder and classifier components to optimize classification performance.
Table \ref{res_bench} presents the results across folds, reported as the mean and standard deviation. Despite the inherent reconstruction loss associated with VQ-VAE approaches, our method achieved performance comparable to or superior to established baselines. 
These representations, with appropriate fine-tuning, demonstrated better generalizability.
This enhanced performance can be attributed to the similarity of our method to metric learning, where input features are systematically optimized to align with selected codebook vectors. The success of this approach provides empirical support for the feasibility of using metastability frameworks in brain dynamics analysis.

  

\begin{figure}[t]
    \centering
    \includegraphics[width=0.9\textwidth ]{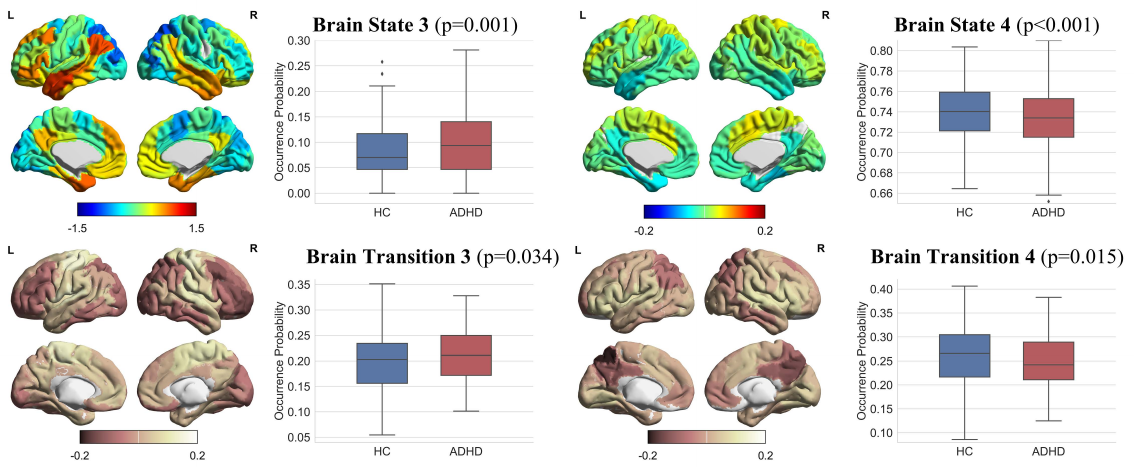}
    \caption{Averaged brain activation maps of brain function states and brain transitions with corresponding occurence probability in significant differences (p-value <0.05) between ADHD and HC groups on the ADHD-200 dataset. The differences remain significant  after FDR correction, except for transition state 3.}
    \label{explain}
\end{figure}

\textbf{4. Brain state and transition quantization analysis.}
We conducted a statistical analysis of brain state and transition occurrence probabilities between groups on the ADHD-200 dataset. Figure \ref{explain} presents the results of comparisons between the groups, highlighting significant differences (p < 0.05). Our analysis revealed that patients exhibited a higher probability of occurrence in brain state 3 and transition 3, while showing a lower probability in brain state 4 and transition 4.
In particular, brain state 4 is associated with reduced activation of brain function. This suggests that patients with ADHD are more likely to experience activated brain function states, aligning with the hyperactivity symptoms commonly observed in ADHD \cite{weyandt2013neuroimaging,yang2024hypercomplex}. Furthermore, significant differences were observed in other brain states and transitions, particularly in the prefrontal and parietal regions, consistent with previous studies on ADHD \cite{yang2023deep,schneider2010impairment}. These findings underscore the feasibility and superiority of our method in dynamic quantization, offering valuable insights into metastability and disease analysis.

\section{Discussion and Conclusion}
In this study, we proposed the HST for brain dynamics quantization, enabling hierarchical quantization of brain states and transitions. We enhanced the metastability quantization of the model by implementing a refined cluster VQ-VAE to improve codebook efficiency and reconstruction performance. Our evaluations show that the HST, with hierarchical quantization, facilitates superior reconstruction and downstream diagnostic performance. Additionally, the HST has the potential to contribute to the development of a brain dynamics foundation model with self-supervised learning, particularly through next-token prediction.

Future work will focus on benchmarking our approach against a broader range of brain state quantization methods and scaling the framework toward large-scale brain dynamics modeling across diverse populations.



\begin{credits}
  \subsubsection{\ackname}
  This work was supported by the German Research Foundation (DFG) Emmy Noether Program (513851350, TW) and the BMBF/DLR project FEDORA (01EQ2403G, TW). We acknowledge the affiliation with the International Max Planck Research School for Intelligent Systems (IMPRS-IS) and the computational resources provided by the de.NBI Cloud, part of the German Network for Bioinformatics Infrastructure (de.NBI).
  
  \subsubsection{\discintname}
  The authors declare that they have no known competing financial interests or personal relationships that could have appeared to influence the work reported in this paper.
  
\end{credits}
  
  \bibliographystyle{splncs04}
  \bibliography{LatexSource-0118}
  \end{sloppypar}
  
  \end{document}